# Accelerated Materials Discovery through Cost-Aware Bayesian Optimization of Real-World Indentation Workflows

Vivek Chawla[1], Stephen Puplampu[1], Haochen Zhu[2], Philip D. Rack[2], Dayakar Penumadu[1], Sergei Kalinin[2]

[1] Department of Civil and Environmental Engineering, University of Tennessee, Knoxville, TN 37996, USA
[2] Department of Materials Science and Engineering, University of Tennessee, Knoxville, TN 37996, USA

**Abstract**

Accelerating the discovery of mechanical properties in combinatorial materials requires autonomous experimentation that accounts for both instrument behavior and experimental cost. Here, an automated nanoindentation (AE-NI) framework is developed and validated for adaptive mechanical mapping of combinatorial thin-film libraries. The method integrates heteroskedastic Gaussian-process modeling with cost-aware Bayesian optimization to dynamically select indentation locations and hold times, minimizing total testing time while preserving measurement accuracy. A detailed emulator and cost model capture the intrinsic penalties associated with lateral motion, drift stabilization, and reconfiguration-factors often neglected in conventional active-learning approaches. To prevent kernel-length-scale collapse caused by disparate time scales, a hierarchical meta-testing workflow combining local grid and global exploration is introduced. Implementation of the workflow is shown on a experimental Ta-Ti-Hf-Zr thin-film library. The proposed framework achieves nearly a thirty-fold improvement in property-mapping efficiency relative to grid-based indentation, demonstrating that incorporating cost and drift models into probabilistic planning substantially improves performance. This study establishes a generalizable strategy for optimizing experimental workflows in autonomous materials characterization and can be extended to other high-precision, drift-limited instruments.

**Introduction**

Materials discovery is a cornerstone of technological advancement. As the demand for performance, functionality, and sustainability grows, so too does the need for new processing techniques and compositions that satisfy the growing needs. Over the last twenty years, considerable effort has been oriented towards theory-based materials prediction[1,2], with companies like Google[3], Meta[4], and Microsoft[5] joining the effort recently. However, the advancements in theory belied the need for experimental realization of theoretical predictions and associated need to accelerate experimental discovery cycles. This led to the wave of interest in the automated synthesis of functional materials via microfluidics[6] and pipetting robotics[7–10], and rapid adaptation of additive manufacturing to the combinatorial discovery.

It is important to note that this wave of interest in high throughput synthesis is but the most recent one. A notable earlier wave emerged in the late 1990s and early 2000s, when advances in thin-film deposition methods, particularly pulsed laser deposition (PLD)[11,12], enabled the creation of large-area composition spreads. The primary lesson of these combinatorial winters was that one of the primary challenges associated with material discovery is not only the generation of candidate materials but also their effective characterization. To keep up with the pace of synthesis, characterization methods must also be altered for speed and precision.

In the realm of mechanical characterization, nanoindentation has emerged as a particularly powerful tool for evaluating combinatorial samples. Originally used for hardness testing and reduced modulus[13–15] evaluation, nanoindentation has evolved into a comprehensive method capable of probing elastic-plastic behavior[16–19], creep[20–24], fracture toughness[25–29], strain rate effects[30–32], and viscoelastic behavior[33,34]. More recent developments have enabled the extraction of full plasticity parameters, including yield strength and ultimate tensile strength[35].

Here, we develop the cost aware framework for active learning in nanoindentation applied to rapid mechanical property discovery in combinatorial libraries using cost-aware Bayesian optimization (BO) with the (structured) Gaussian Process surrogate models. Here, the ML-enabled nanoindentation or automated nanoindentation (AE NI) system, dynamically plans the experiment, deciding the location of subsequent indentation based on the combination of mean predictions, uncertainty and time required for these measurements. The unique aspect of nanoindentation is that it presents intricate measurement costs associated with the manufacturer-specified engineering controls in the system. Here, we calibrate these measurement costs and derive the universal cost function structure for BO. By coupling the Gaussian Process (GP) model with different noise priors (classical and heteroskedastic) with a custom acquisition function that accounts for mean, uncertainty, and the cost, a cost aware decision-making algorithm is defined. We report the model simulated workflows and experimental implementation on the model KLA nanoindenter systems, overall enabling the guidelines for the discovery.

This study advances the concept of automated, adaptive mechanical characterization by operationalizing a cost-aware Bayesian optimization framework for nanoindentation. The proposed workflow unifies probabilistic modeling, emulator-based cost calibration, and experiment planning within a single decision-making loop that accounts for both measurement uncertainty and instrument-imposed overheads such as reconfiguration, positioning, and drift stabilization. To address the kernel-length-scale collapse arising from measurements across different temporal regimes, a hierarchical meta-testing strategy is introduced that combines local grid exploration with global optimization. The resulting AE NI system dynamically balances exploration and efficiency, reducing total experimental time by more than an order of magnitude compared to conventional grid mapping, while maintaining accuracy in reconstructed property fields. Beyond a proof of concept, the framework generalizes to diverse materials systems and provides a practical foundation for intelligent automation in high-throughput mechanical characterization.

## *I. Emulator for Thin films*

Figure 1 illustrates the envisioned operation of the automated experimental nanoindentation (AE-NI) framework. In this approach, a surrogate regressor continuously learns from previously measured indentation data to predict the evolving property landscape and its associated uncertainty. Based on these predictions, the controller identifies the next most informative measurement point and commands the nanoindenter to execute it, forming a closed loop between computation and experiment. This adaptive cycle enables efficient exploration of thin-film libraries by prioritizing measurements that maximize information gain while minimizing time and repositioning costs.

To illustrate the workflows for the automated structural property discovery via ML-enabled nanoindentation and given that Bayesian Optimization to our knowledge has not been used for automated nanoindentation before, we first illustrate the emulator for workflow. In addition to the illustration purposes, such emulators can significantly facilitate the initial hyperparameter tuning for the ML algorithms, required given the high cost of experimental measurements compared to the theoretical workflows.

Here, the simulated combinatorial library, representative of a ternary alloy system, is created by diagonally varying the composition of elements A, B and C (Supplemental information Figure S1) onto a two-dimensional spatial domain thereby emulating a realistic deposition pattern often seen in combinatorial synthesis. The ground-truth response was generated using a synthetic model that incorporates multiple compositional effects, including baseline mixture behavior, pairwise interactions, and higher-order concentration-dependent terms. The resulting hardness was the sum of the three mechanisms as shown in Equation 1. More details on the choice of ground truth can be found in Supplemental information. The resulting hardness variation can be found in supplemental information (Figure S1).

$$H = \sum_{i=A}^{C} c_i . H_i + \sum_{i,j=A}^{C} \sigma_{ij} . c_i c_j + k * c_C^2 * (1 - c_C) \qquad \text{-(1)}$$

Where, $H$ is the overall hardness. $H_i$ are baseline elemental contributions, $c_i$ is the concentration of element i, $\sigma_{ij}$ are the interaction coefficients, and k modulates the nonlinear concentration dependent term.

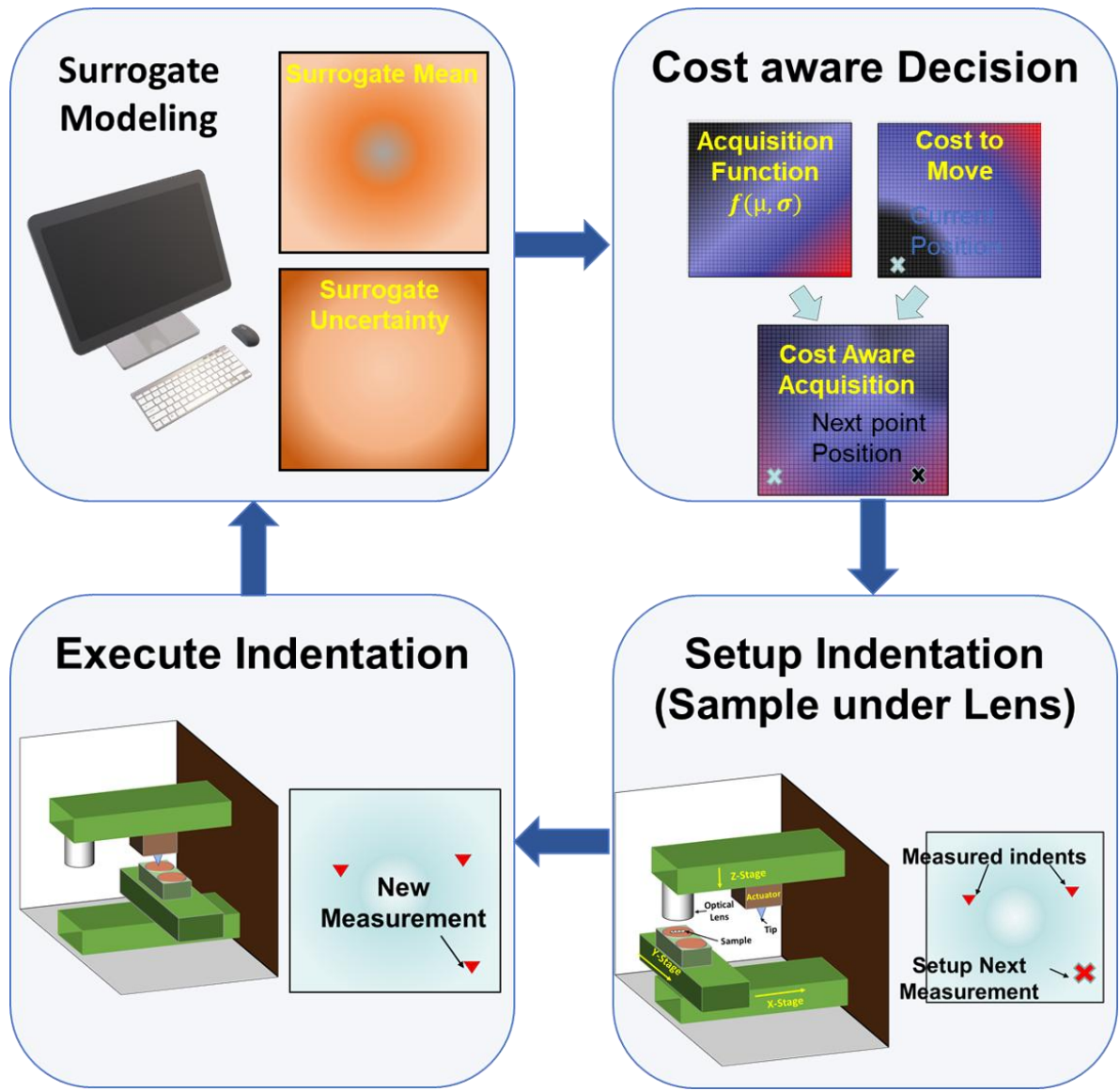

*Figure 1.* *Schematic of the automated experimental nanoindentation (AE-NI) framework. A surrogate model continuously updates its mean and uncertainty from measured indents, guiding the selection of the next measurement based on uncertainty and cost. The controller converts model decisions into stage movements, closing the loop between data acquisition, model update, and physical indentation for adaptive mechanical mapping.*

To enable automated experiment, we use Bayesian Optimization (BO) based on surrogate Gaussian Process Based model. BO is a probabilistic framework designed to efficiently optimize or explore functions that are expensive to evaluate, often referred to as black box functions. In the context of automated experiments, BO has emerged as a powerful strategy to reduce the overall cost of discovering or optimization of the black box function. In the current emulator, the black-box function corresponds to the unknown spatial variation of hardness across the X-Y space. BO helps navigate the trade-off between exploration (sampling the regions of high uncertainties) and exploitation (sampling regions that based on current analysis are most likely to yield the optimal values) to achieve experimental objective. It is important to note that key aspect of BO compared to other decision-making frameworks (reinforcement learning, dynamic programming, etc.)[36,37] is that the decisions are myopic, ideally matching the exploration of combinatorial libraries where the results of measurements in a certain location do not depend on previous measurements elsewhere.

To achieve this objective, Bayesian Optimization requires a surrogate model that relates the input feature space (X, Y) to the target property (hardness). In this study, a Gaussian Process (GP) as a surrogate model is used. GPs are probabilistic models that assume the value at any given location is statistically correlated with values at nearby locations. Mathematically, a Gaussian Process (GP) is defined in Equation 2. Specifically, nearby points have stronger influence, while the influence of distant points diminishes with increasing distance. The spatial correlation is captured through a covariance matrix, whose entries are determined by a kernel function. For example, the Radial Basis Function (RBF) kernel (Equation 3), scales this correlation based on a kernel length ($l$). A smaller kernel length implies rapid function variation over short distance or noise effects. Similarly, a large length scale indicates slow variation over large distances.

$$f(x) \sim GP(\mu(x), k(x, x'))  \quad -(2)$$

$$k(x_i, x_j) = \exp\left(-\frac{||x_i - x_j||^2}{2l^2}\right) \quad -(3)$$

During GP training, both the mean and kernel parameters are optimized to best represent the underlying black-box function. The key advantage of GPs is that they provide not only a prediction at each point but also an estimate of the uncertainty associated with that prediction.

This enables strategic decision-making in Bayesian Optimization. Depending upon the objective, whether to explore uncertain regions or exploit the regions with high predicted values, the next sampling point can be chosen. This BO decision-making process is governed by the acquisition function. The acquisition function balances the exploration and exploitation and guides the selection of the next query point. Some of the most commonly used acquisition functions include Upper Confidence Bound (UCB), Expected Improvement (EI), and Uncertainty Estimation (UE). UCB favors points where either the predicted mean is high or the uncertainty is large. EI quantifies the expected amount by which the new point will improve over the current best which inherently balances the mean and variance by giving more value to points that are likely to improve performance. UE acquisition strategy simply choses the points of highest uncertainty. Mathematically the UCB, EI and UE are defined in equation 4, 5 and 6 respectively.

$$UCB(x) = \mu(x) + \beta^{1/2} \cdot \sigma(x), \beta \text{ controls the trade off} \quad -(4)$$

$$EI(x) = E[\max(0, f(x) - f^+)], f^+ \text{ is best observed so far} \quad -(5)$$

$$UE(x) = \sigma(x) \quad -(6)$$

From the perspective of this study, there are two possible choices for the feature space used in the surrogate model: one can use either the spatial coordinates (X, Y) or the compositional variables (fractions of A, B and C) as input features. From both an experimental and emulator perspective, a gradual and continuous variation in composition across the X-Y space is expected. As a result, the system effectively has only two independent variables, and therefore the X-Y coordinates are used as input features.

Note that BO can be based on more complex surrogate models, as long as they allow interpolation over combinatorial space and either allow for prediction and uncertainty, or allow for multiple probabilistic function prediction (Thompson sampling). Therefore, the workflow proposed here can be elementarily updated to include such alternative models.

Another important aspect of BO is that typically in theoretical implementations the costs of measurements everywhere in the parameter space is assumed to be constant. Practically however the cost structure (e.g., time of measurement, computational cost, etc.) can be readily incorporated into the acquisition function. In nanoindentation measurements, the measurement cost can be very complex and is determined by the engineering controls established by the manufacturers, and is discussed below.

## II. Experimental implementation

Implementation of ML-driven active learning in combinatorial nanoindentation requires careful consideration of actual experimental workflows. In this work, the analysis is specific to a KLA nanoindenter. For different instruments from the same manufacturer, the procedure can be reproduced through calibration, while for instruments from other manufacturers the overall workflow of automated CSM indentation remains the same but individual details may differ. For instance, grid positioning and setup can vary between systems, as can testing details such as the method of surface detection and the timing and duration of drift evaluation. Nevertheless, the methodology is analogous. All results here are presented for the continuous stiffness measurement (CSM) technique, which is the standard approach for measuring hardness and modulus[14], although the same ideas can be extended to other testing modes such as fracture toughness or displacement-controlled experiments. Inevitably, the individual details of implementation will change. Key considerations are the cost of stage movements, drift effects, and manufacturer-imposed constraints, which together form the basis for benchmarking and uncertainty quantification.

### II.1. Automated Nanoindentation Workflow

A KLA iMicro Nanoindenter system is used in this current study. In our previous work[38], an automated nanoindentation framework was developed that enables the direct control of the indenter through a Python environment. The automated system allows users to specify indentation locations based on X-Y coordinates, the indenters optical image, or an external optical image. For this work, the focus is exclusively on X-Y coordinates-based indentation, as the emulator exhibits continuous variation of composition across the physical X-Y space with no distinct microstructural features.

The nanoindentation workflow follows a standardized sequence. First, samples are mounted on a sample holder and loaded within the nanoindenter. The holder then moves from the 'safe-position' to a location beneath the optical lens via X-Y movement, while the Z-stage is subsequently adjusted to bring sample into focus. This step is called the initialization step and it remains the same across both manual and automated nanoindentation workflows. Following the initialization, indentation grids (series of indents that follow a consistent spacing in x and y) are defined. The definition of grids consists of moving to the position of the grid underneath the optical lens, setting up the grid (setup step) and repeating the procedure until all the grids are defined. Subsequently, the indentation procedure is carried out. The indentation procedure follows a structured sequence: the Z-stage rises (frame with indenter and optical lens moving away from the sample) and the X-Y stage shifts to the location of first indent in first grid under the indenter (lift step), the Z-stage then lowers to detect the surface contact (engage step). Subsequently, the system waits for stabilization of the drift (drift settlement step) before proceeding to test each indent within the grid (measurement step). Henceforth, the combined sequence of Z-stage upward movement, X-Y stage movement, surface engagement, and drift stabilization is referred to as reconfiguration. After finishing a grid, the reconfiguration happens as the sample moves between consecutive grids. Once the final grid is completed, the Z-stage rises and the sample moves from underneath the indenter back to the optical lens.

### II.2. Modeling Experimental Cost

To implement active learning on nanoindenter, it is first essential to quantify the cost associated with each action the nanoindenter performs. Here, we build a detailed cost model that captures the real-world time penalties tied to each step in the indentation workflow.

The time associated with lift step ($t_{step}$), engage step ($t_{engage}$), and comeback step ($t_{comeback}$) was recorded to be approximately 30s, 65s, and 35s, respectively. In addition to the above, there are costs that are variable during indentation. These costs include drift stabilization time ($t_{drift}$), Measurement time ($t_{measurement}$) and movement time ($t_{move}$). The average drift is set to 300s ($t_{drift}$= 300 s), although in practice it can vary and can be bounded. The $t_{measurement}$ typically depends upon the type of test as well as the target depth. $t_{measurement} = 120$s is assumed based on the target indentation depth of 200 nm and Continuous Stiffness measurement (CSM) based indentation with strain rate of 0.2. Lastly, the $t_{move}$ is a function of Euclidian distance between indentation locations. Therefore, $t_{move}$ is modeled by recording the time required to perform stage movements ranging between 2 and 50000 µm.

The movement time is then fit using a piecewise function consisting of an exponential model for short range movements and a linear model for long range movements. This functional choice is based on the behavior of stepper motor-controlled stages, which often use PID control systems that behave differently depending on movement scale. Figure 2a shows the cost of movement from one location to another. Figure 2b shows the function during indentation. As can be seen there is a jump of ~400 seconds in time required to move from one location to another if the distance between the two location is greater than 1000 µm. This is an important manufacturer-imposed safety constraint to prevent the indenter tip from accidently striking the surface during extended movements. Therefore, to move such large distances, a reconfiguration penalty is applied i.e., the system first raises the Z-stage ($t_{rise}$), then the sample is moved laterally to new location ($t_{move}$) followed by lowering the Z-stage ($t_{engage}$) and drift stabilization ($t_{drift}$). To demonstrate the impact of this constraint, two indentation grids are modeled (Figure 2c). In the first grid, three indents are placed with the first two spaced by 100 µm, and the third indent is located 990 µm away from the second indent. In the second grid, the same positions are used except the third indent is placed 1000um from the second indent- just crossing the reconfiguration threshold. Figure 2d shows the resulting total time per configuration showcasing the ~400s discrepancy in the second case.

Figures 2e and 2f illustrate the full sequence of steps during indentation for the two cases. In the no-reconfiguration case (Figure 2e), once the grid setup, the z-stage rises and sample moves underneath the indenter, and waits for the initial drift stabilization to be complete. All subsequent indents are then executed sequentially within the same region. Only the indenter tip and x-y stage move between points, leading to minimal overhead and rapid measurement cycles. In contrast, when the displacement exceeds the 1000 µm threshold (Figure 2f), the system performs a complete reconfiguration sequence before the third indentation. This additional sequence introduces a fixed time penalty (~400s) per reconfiguration event, significantly increasing the total experiment duration. Together, these steps emphasize that the cost function captures not only lateral motion time but also the real operational overhead imposed by the instrument's safety and drift-control protocols.

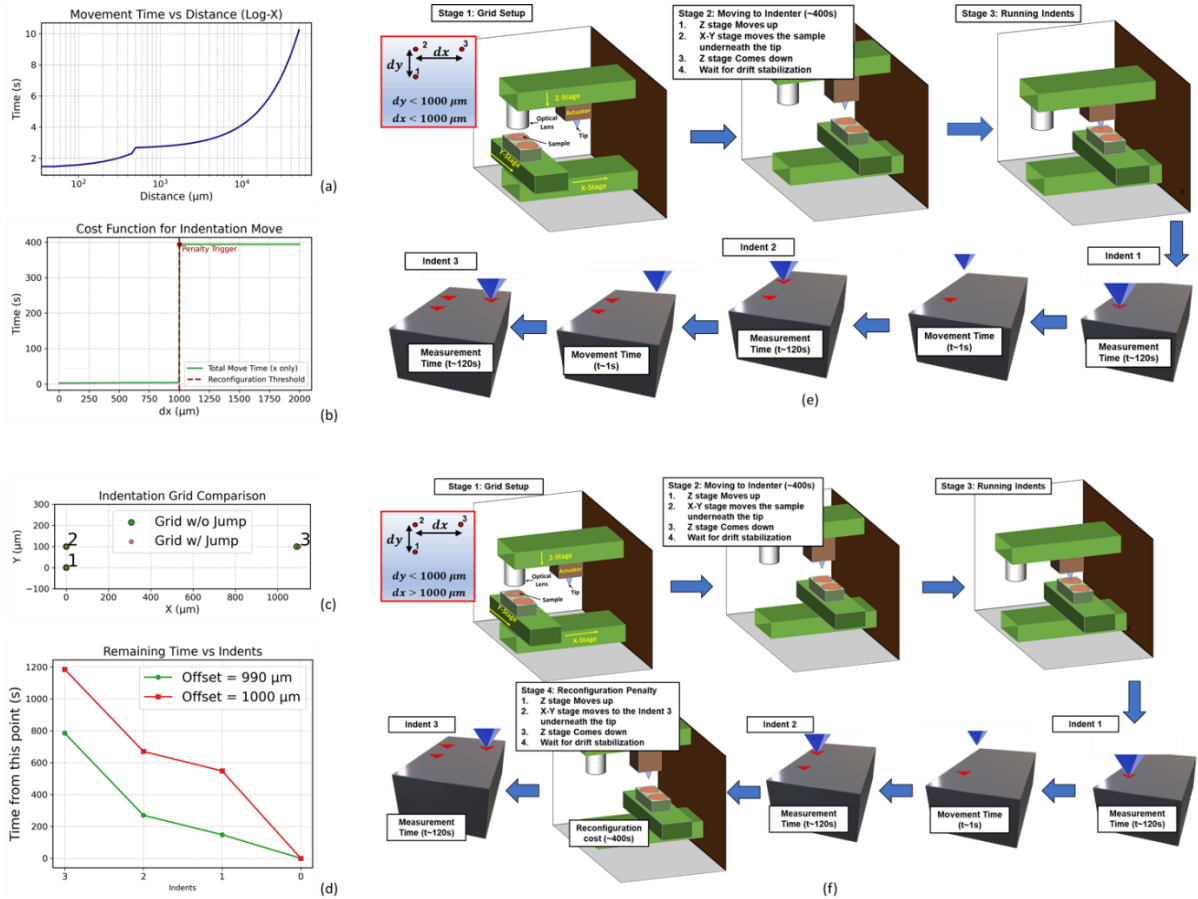

*Figure 2. Movement-based cost model for adaptive indentation. (a, b) Movement time and corresponding cost function showing a 400 s reconfiguration penalty for travel distances exceeding 1000 μm during indentation. (c–f) Example grids and execution sequences demonstrating how the penalty alters total measurement time: when displacement exceeds the threshold, full Z-stage repositioning and drift stabilization are triggered before resuming indentation in f (travel distance > 1000 μm) while only indenter moves in e (travel distance < 1000 μm)*

## II.3. Building baselines:
### II.3.1. Grid Search

With the cost model in place, the next step is to establish a baseline experimental cost to allow for benchmarking of BO based active learning vs. classical approach. In conventional nanoindentation workflows, the domain is often explored using a grid search approach. To mimic this common practice, a 9 * 9 grid search over the emulator domain is performed, where each grid contains 5 * 5 indentations, resulting in a total of 2025 indents. Using the previously defined cost model, this exhaustive search corresponds to a total testing time of approximately 2,863,900s (~80 hours). Figure 3a shows the grid search pattern on the true variation of hardness, while Figures 3b and 3c display the ground truth hardness map and the reconstructed estimate, respectively. Although the variation in mean absolute percentage error (MAPE) converges only marginally faster for the regression-based adaptive approach compared to the grid search (Figure 3d), prior studies have demonstrated that active learning methods can improve convergence rates by a factor of two to three under more complex conditions. The relatively modest gain observed here likely stems from the simplicity of the emulator's ground truth surface. Nevertheless, despite the extensive number of indents and significant time investment, the grid search fails to identify the true maximum hardness located near the bottom-left region of the domain (Figure 3e). While targeted edge indentations could, in principle, reveal this feature, such decisions require prior knowledge of the ground truth. Consequently, any sharp gradient or localized phase transition would remain undetected within a uniform grid scheme. Adaptive exploration strategies, by contrast, incorporate feedback from measured data, enabling dynamic refinement of sampling density in regions of high variability. This capability transforms data collection from a passive mapping exercise into an information-driven process, thereby reducing experimental cost while improving the likelihood of capturing critical microstructural or compositional transitions.

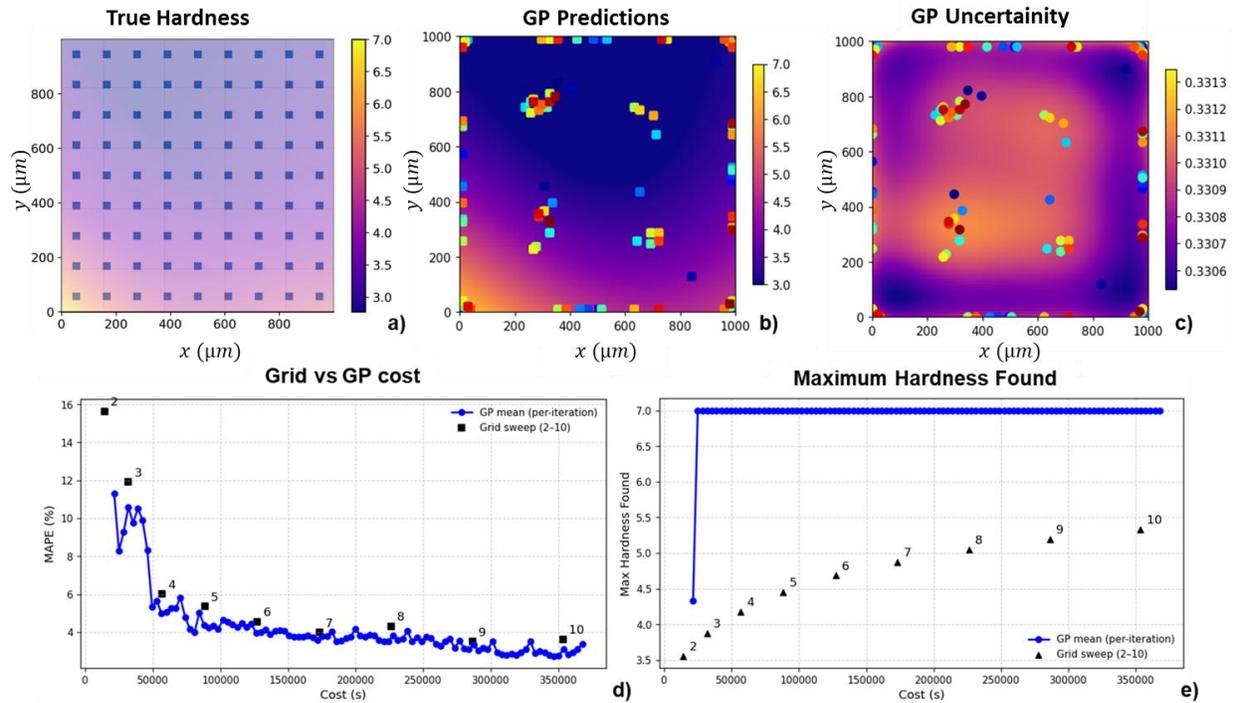

*Figure 3. Comparison of grid search and GP exploration. (a) True hardness map with a 9×9 grid sampling pattern. (b, c) Adaptive regressor predictions showing measured points (recent in red, early in blue) and corresponding uncertainty. (d, e) Mean absolute percentage error and maximum hardness found versus total cost, illustrating faster discovery and convergence of the adaptive method compared to grid-based search.*

### II.3.2. Measurement cost models

With the cost model established, we analyze the measurement bottlenecks, i.e., the stages where most of the experimental time is spent during testing. For a given grid, total time ($t_{total}$) = $t_{move}$ + $t_{Engage}$ + $t_{drift}$ + $t_{Measurement}$ + $t_{lift}$. An obvious next step is to evaluate the percentage of time spent on actual indentation measurements. This percentage would depend upon testing parameters as well as grid size once the assumption is made on $t_{drift}$ and $t_{Measurement}$. The distribution of this cost is shown in Figure 4a and 4b (for cases of grid spacing being below and above 1000 µm). For the single indent grids, ~60% of the time is spent in drift calibration, making the test highly inefficient. In contrast, for larger grids, the majority of the time is spent on measurements. Of course, the total time per grid is also higher for larger grids. If the grid spacing is above 1000 µm (Figure 4b), most of the time is spent on reconfiguration cost. As most time is spent on measurements for higher grid size, it is useful to break down the measurement duration itself. Following the Oliver-Pharr method and default Continuous Stiffness Measurement (CSM) settings, ~ 30s is spent on detecting surface, ~40s for measurement with a ~2 s hold (crucial for elastic unloading for viscoelastic materials), ~ 80s is allocated to post-hold drift monitoring, which is then used to correct the displacement data. Thus, nearly 50 % of the measurement time is used just to evaluate drift. This is notable because a significant time is spent allowing drift to settle before testing begins (~300s). Given this, there is a strong case for adaptive drift monitoring strategies, especially in exploratory workflows where the goal is to understand property variations. Since drift is often caused by external thermal or mechanical fluctuations, adapting its measurement dynamically can potentially reduce time without significantly affecting data quality as drift is often caused by an external event.

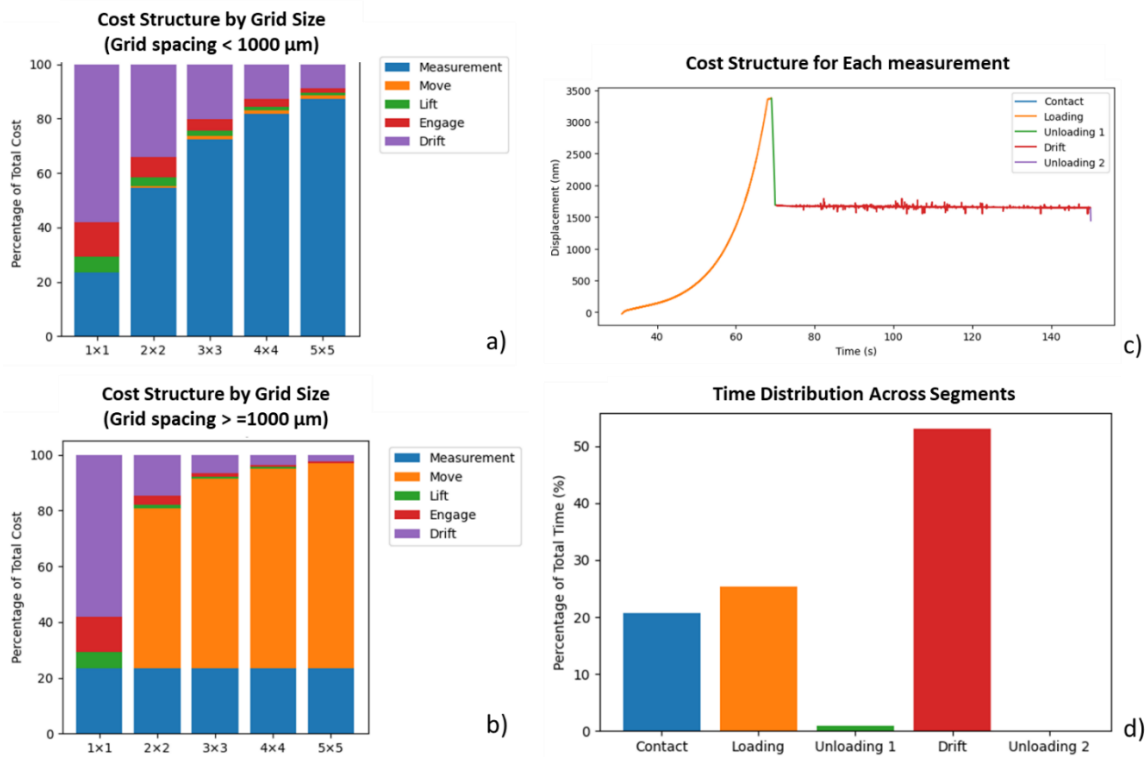

*Figure 4.* Cost structure and measurement characteristics. (a) cost distribution for grid spacing <1000 μm. (b) cost distribution for grid spacing > 1000 μm. (c) measurement time distribution for fused silica indents with maximum target load. (d) percentage time spent during measurement

### II.2.3. Lateral drift:

An important fact in any experimental measurements is drift, i.e., deviation between intended measurement location with the actual one. In ML-enabled experimentation, the drift role is twofold. The negative aspect is the loss of positional precision, affecting the surrogate models that are traditionally much more sensitive to the uncertainties in parameter space then in the measured function. The positive is that accelerated discovery compared to the grid-based operations reduces the deleterious effects of drift. We note that in principle drift can be accounted for post-acquisition (simply by mapping the locations of indentation sites) or dynamically via the use of optical fiducials, but defer these studies to the future. Here, we explore the effects of drift on model performance

*II.2.4. Adaptive Drift sampling*

Vertical drift in nanoindentation experiments is often caused by thermal relaxation, instrument instability and sample movement. Drift does not result in noisy experiments but causes a bias in the measured depth as given by Equation 7 (where DR is the measured drift rate) which in turn results in the bias in the area function and subsequently, the hardness (Equation 8-10). Assuming the realistic values of DR of 0.1 nm/s and loading segment of 30 s for an indent depth of 3000 nm, results in an error of less than 0.1 percent in hardness. Therefore, a long drift hold (~80 s) typically used in traditional workflows is not necessary. While effecting in eliminating bias, this hold is time consuming. This approach assumes drift is always significant, which may not hold in practice, especially in short or localized experiments.

$$h_{measured} = h_{indent} + t_{measurement} * DR \quad\quad -(7)$$

Assuming,
$$h_{measured} = h_{contact}, \text{i.e., } \epsilon = 0 \quad\quad -(8)$$

$$H = \frac{P}{A_c} \quad\quad -(9)$$

Where, $h_{measured}$, $h_{indent}$, and $h_{contact}$ are the measured depth, total indentation depth, and the contact depth. $\epsilon$ is the Oliver-Pharr[14] epsilon that accounts for pile-up and sink in effects. $H$ is the Hardness, $P$ is the load, and $A_c$ is the contact area.

Therefore, a more adaptive strategy is adopted that balances drift correction with experimental efficiency. We use an adaptive holding scheme with two hold durations: 5seconds and 80seconds. The short 5s hold is used by default, while the 80s hold is activated when the estimated drift exceeds a defined threshold and remains until the drift stabilizes. Two drift scenarios are analyzed. In the first, the drift-induced error is small (0.3) and quickly recovers. In the second, the drift error is large (~1) and decays slowly, as shown in Figure 5a. Figures 5c–5e show the GP predictions, uncertainty, and deviation from ground truth for the bad drift case. In the favorable drift case, the adaptive hold strategy preserves model accuracy while reducing total measurement time by up to a factor of about 2 compared to a fixed 80s hold. In the high-drift case, the model automatically increases the hold duration, yielding smaller time savings but maintaining reliable convergence. The exact improvement factor depends on drift magnitude and parameters, but overall, the results demonstrate that adaptive holds can yield significant time gains while retaining predictive robustness (Figure 5b).

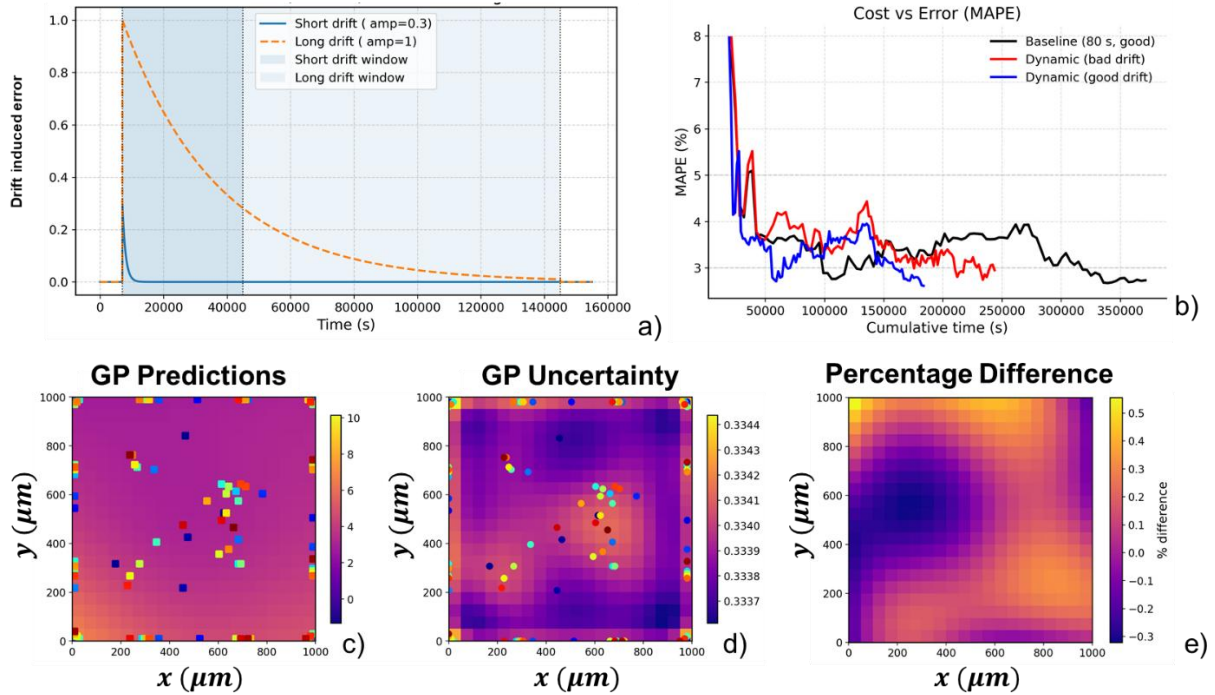

*Figure 5. Effect of drift-aware adaptive hold selection on mapping accuracy. (a) Short- and long-drift models showing exponential decay of drift rate over time. (b) Mean absolute percentage error (MAPE) versus cumulative time for baseline (fixed 80 s hold) and dynamic hold strategies under good and poor drift conditions. (c–e) Corresponding regressor predictions, uncertainty, and percentage difference maps, demonstrating that drift-adaptive scheduling maintains accuracy while reducing total experiment time.*

## III. Building search strategies

In principle, the analysis of the cost function enables construction of a purely Bayesian optimization (BO) workflow, where the Gaussian Process (GP) surrogate suggests the next measurement location (based on the policy) under cost constraints. However, in practice such an approach often demonstrates instabilities, e.g., the collapse of kernel length scales. This happens due to the noise in the surrogate function coupled with large time differences between repeated local measurements vs. measurements that require large stage movements. In principle, this can be corrected by the non-Gaussian (e.g., student t-distribution) based noise models, but that can lead to introduction of a lot of additional hyperparameters.

While these are guaranteed to self-correct over long measurement times, practical considerations of AE NI can benefit from more robust schemes. One such approach is human in the loop[39,40]; another is introduction of more robust policies. Here, we do it by analyzing grid measurement v. local measurement grid.

### III.3.1. Adaptive Grid Sizing

Often in indentation testing, a $n_x * n_y$ grid is conducted to get measurement at a 'single' location. These indents act as multiple tests at single location to understand the noise associated with the measurement. Therefore, the numbers of indents are often decided by a probabilistic model, like the student-t distribution, such that the confidence in the measurement is high. Often,

$n_x = n_y = 5$ are set, resulting in a cost of ~3500 s per grid. However, the grid size is predetermined and does not adapt to the underlying variation within the sample.

In nanoindentation of combinatorial samples, spatially varying noise is not uncommon. By definition, combinatorial libraries exhibit compositional gradients, and each composition can lead not only to different target properties (e.g., hardness) but also to different noise levels (arising from processing steps, layer adhesion, and sample curvature). In such cases, a heteroskedastic Gaussian Process (GP) is more appropriate as it not only adapts to the underlying trend but also learns a spatially varying noise function, treating noise as a variable rather than a fixed quantity. In heteroskedastic GP, two GPs are trained simultaneously: one modeling the mean of the data and the other models the input-dependent noise. When the noise is truly gaussian and randomly distributed, the heteroskedastic GP naturally reduces to the behavior of standard (homoscedastic) GP, effectively recovering the noiseless case.

To realize this, firstly noise is added to the nanoindentation measurements. Figure 6a-c shows the ground truth of the emulator, the noise added to the measurements, and the measurements after the added noise. For the purpose of this work, the maximum noise was kept low compared to the maximum measured hardness (~1.4 percent). Given the measurements, we actively train the heteroskedastic GP and compute the expected noise at any candidate point. Given a target standard error of the mean (target SEM), denoted by $\epsilon$ and given by the user, the required number of indents can be determined at that location using Equation 10. To keep the grid size practical and avoid extremes, clipping and rounding is applied to resulting grid size using equation 11. This means the system will adaptively choose between 2*2 to 5*5 grid size, depending upon local noise levels.

$$n_{req}(x, y) = \left(\frac{\sigma_{ext}(x, y)}{\epsilon}\right)^2 \qquad \text{-(10)}$$

$$n_{grid} = \min\left(\max\left(\sqrt{n_{req}}, 2\right), 5\right) \qquad \text{-(11)}$$

$where, n_{req}$ is the minimum number of grid required, $\sigma_{ext}(x, y)$ is the estimated noise at location $(x, y)$, $\epsilon$ is the target SEM, and $n_{grid}$ is the chosen grid size. Figure 6e-f shows the predicted mean, and the expected noise. In this case, even though the model was not exactly able to recreate the noise model (perhaps because the noise itself was kept low), the target SEM criterion was sufficient to guide intelligent grid resizing. The evolution of grid size with iteration is shown in Figure 6d. Furthermore, the adaptive approach achieves approximately a fourfold reduction in total time compared to the fixed-grid case, while reaching a comparable error level (see the blue and black curves in Figure 6j).

### *III.2. Meta Grid*

As further optimization of automated indentation workflow is carried out, there is a clear bottleneck: reconfiguration cost. As showed in Figure 4a, particularly for smaller grids (which become dominant in low noise regions), a significant portion of the total time is spent transitioning between grids rather than performing indentations. To address this bottleneck, a strategy is introduced that groups multiple local grids, thereby forming a meta grid, within a single global "safe zone". For instance, considering a case where, based on the noise estimate from heteroskedastic GP, a 2 * 2 grid is required at a global location of $(g_{x1}, g_{y1})$, a set of neighboring 2*2 grids centered at $(g_{xi}, g_{yi})$, where $i = 1: N$, are predefined. Collectively, these $2*2*(N + 1)$ indents form what is subsequently referred to as the meta grid, with each 2 * 2 set defined as the sub-grid. The next step is to determine the locations $(g_{xi}, g_{yi})$ for the sub-grids. These can be chosen using various strategies. For example, selecting locations of highest uncertainty, combining uncertainty with distance-based weighting or using uniformly spaced points. Regardless of selection strategy, the manufacturer introduced constraint of no consecutive indentation being more than 1000 um apart must be followed to avoid triggering the reconfiguration penalty. Another essential part of the strategy must include defining a safe zone in the Z-direction. In our previous work[38], it has been demonstrated that the topography variation of the sample can be approximated by a linear smooth surface fitted using X, Y, Z coordinates under the microscope (or the indenter). The details are part of the broader automation framework and can be found elsewhere[38]. Once a local Z-gradient is known, a constraint that any movement from the global grid center $(g_{x1}, g_{y1})$ stays within a maximum allowable vertical deviation, i.e., $\Delta Z < \Delta Z_{safe}$ can be enforced. This defines a safe operating region within which sub-grids can be explored

In the current study, the above is shown using a predefined layout in which sub-grids are spaced 500 µm apart, forming a rectangular pattern within the safe zone. The safe zone is approximated based on an assumed linear Z-surface in the emulator. After each sub-grid is completed, the algorithm evaluates whether to proceed to a new global location, which comes with a reconfiguration cost, or to continue within the current safe zone. This decision is made using $\frac{\sigma}{cost}$, where $\sigma$ represents model uncertainty, and $cost$ represents the time penalty of moving. Initially, the algorithm favors global moves, but as uncertainty reduces, it shifts to exploring local sub-grids. Figure 6g illustrates the above algorithm, where black dots denote globally visited points and blue dots represent locally visited points during current iteration. The center of each local grid represents the current global center, and the surrounding blue points form the sub-grid selected by the algorithm for investigation before moving to the next global center. The diagonal region enclosed by the dashed line represents the safe zone defined for the current global center. In this iteration, 8 sub-grids were explored, all without triggering reconfiguration penalties. Figures 6h-6i show the predicted mean hardness, and the predicted noise variation, respectively. The model not only captured the spatial noise variation qualitatively but also achieved nearly a tenfold reduction in total time for a comparable error level (see the red and black curves in Figure 6j). The inset in Figure 6j illustrates the number of local grids executed within the meta-grid framework. Initially, only a single grid was explored; however, as global uncertainty decreased, an increasing number of local grids were adaptively investigated. As noted earlier, caution is warranted since excessive local exploration can lead to kernel length collapse, where the model begins interpreting noise as meaningful data. Therefore, incorporating a human-in-the-loop component may be essential to maintain model stability in those cases.

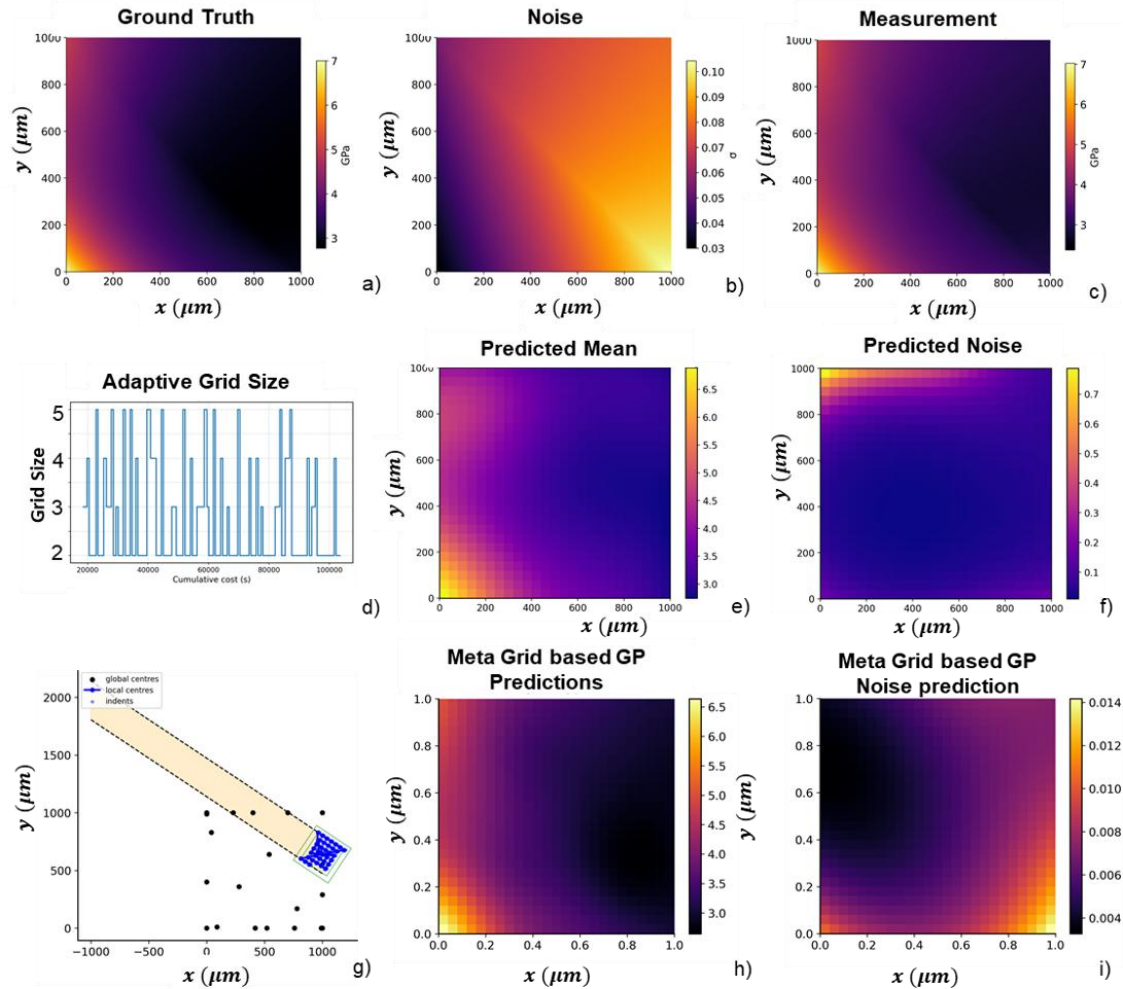
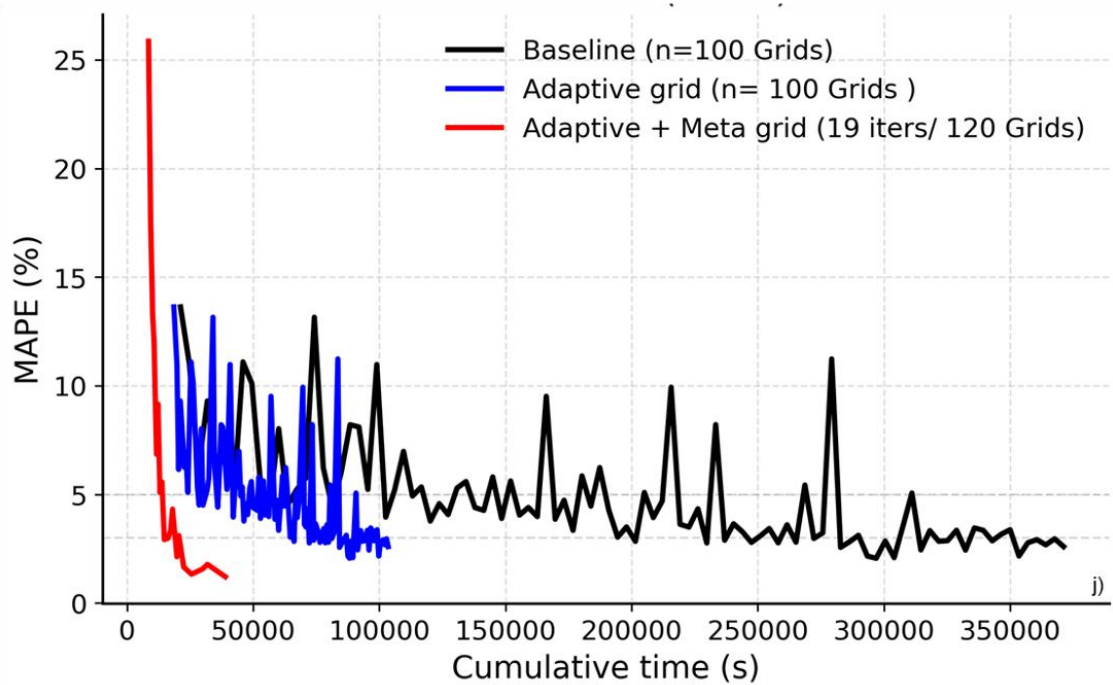

*Figure 6. Adaptive grid sizing and meta-grid strategy under heteroscedastic noise. (a–c) Ground-truth hardness, spatially varying noise field, and corresponding noisy measurements. (d) Evolution of grid size with cumulative cost showing dynamic resizing between 2×2 and 5×5 based on local noise. (e, f) Predicted mean and noise fields from the trained heteroscedastic regressor. (g–h) Meta-grid configuration showing transition from local to global exploration. (i–j) Comparison of mean absolute percentage error versus cumulative time for baseline, adaptive, and adaptive + meta-grid cases, demonstrating a fourfold reduction in total experiment time while maintaining accuracy.*

## IV. Materials discovery via Adaptive exploration:

To experimentally demonstrate the proposed cost-aware adaptive indentation framework, the workflow was implemented on a co-sputtered Ta–Ti–Hf–Zr thin-film combinatorial library (Materials and Methods). This section translates the simulation-based active learning and cost-modeling concepts into a real experimental setting. The film serves as a model quaternary system with continuous composition gradients, allowing direct testing of the framework's ability to adaptively plan indentations under realistic instrument constraints. The thin-film indentation workflow was experimentally automated by combining stage-space parametrization, heteroskedastic Gaussian-process modeling (both mean and noise are fitted with GP with constant mean prior), and cost-aware planning under instrument and drift constraints as described in previous simulations. Four calibrated corners- top-left (TL), top-right (TR), bottom-right (BR), bottom-left (BL)-define an affine map from stage coordinates $(x, y)$ to unit-square coordinates $(u, v)$ using BL as origin. The film is discretized into 2,500 candidate centers (50×50) in stage space (Figure 7a), while modeling and acquisition operate in $(u, v)$. Two initial 5×5 local indentation grids are conducted at random centers with 80s hold. Each indentation grid is collapsed to one robust Hardness measurement for a tested center given by the median of the measurements.

Prior to scheduling any local indentation grids, the "safe band" is computed to avoid topographic violation. Subsequently, the heteroskedastic GP models hardness median in normalized space $(u, v)$ and then reports the mean predictions and uncertainty in physical x-y space. Consequently, uncertainty is used to identify the next global or primary center and grid size respectively. Candidate centers are ranked by uncertainty divided by time cost. The local cost scales with current grid size $(g) * (move + hold)$ while the global cost includes an additional reconfiguration of 520s. Hold-time selection is drift-aware i.e., using the last measured drift rate $(DR)$ and a nominal maximum depth of 100 nm (target indentation depth), the percent area error is $|\frac{h_{meas}^2 - h_{true}^2}{h_{true}^2}| \times 100$ with $h_{true} = h_{meas} - DR \times t_{measurement}$ is determined and subsequently if the error is below 2% the method "hold = 5s" is chosen, otherwise "hold =80s" is chosen.

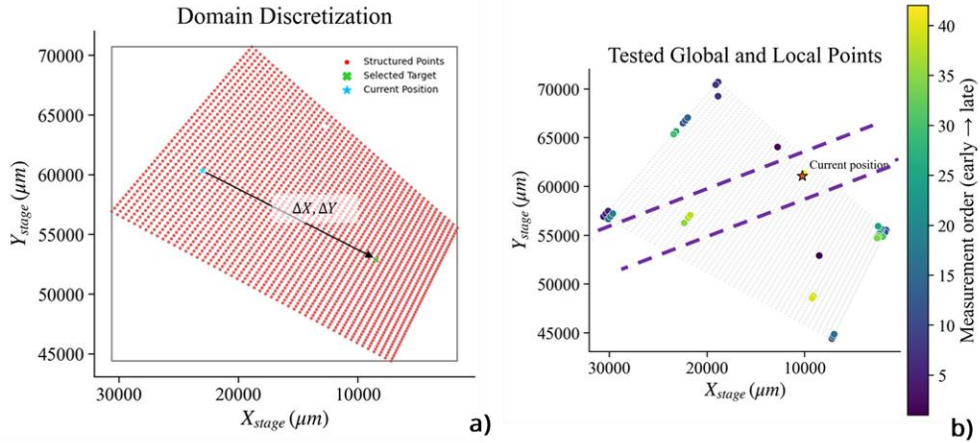

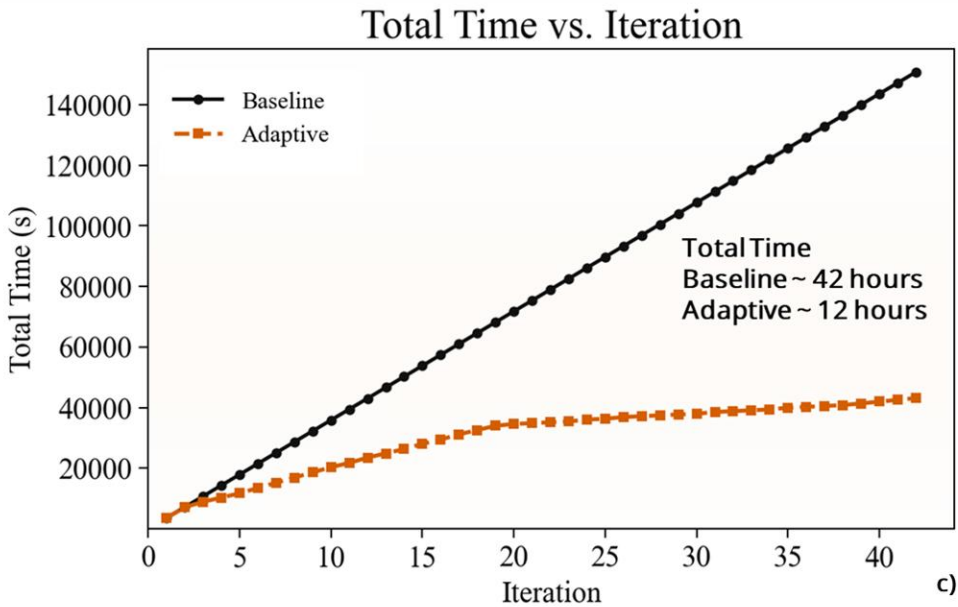

*Figure 7: Illustration of global–local exploration during adaptive indentation. (a) A 50×50 sampling grid, shown with the x-axis reversed to match the indenter stage motion. (b) Example of executed centers (colored by order), the next global candidate (star), and planned local centers.*

After the two initial randomly chosen 5×5 indentation grids, the active-learning loop begins. The early seed points exhibited high measurement noise (higher standard deviation within a grid). In such cases, the Gaussian Process (GP) model tends to remain highly uncertain, causing the exploration to become locally confined and potentially leading to kernel instability ("kernel crash"). To mitigate this, the first four iterations (the exploratory phase of active learning) were restricted to a minimal number of local centers to promote stable kernel formation. Local centers were identified by estimating a planar surface model, $z(x, y) = ax + by + c$, fitted to three measured corner elevations at the top-left (TL), top-right (TR), and bottom-right (BR) points, providing an estimate of the local surface slope and offset. This model was used to define a safe band around each global center, within which indentation locations were permitted. During each global iteration, local centers were queued sequentially before testing. After the first four exploratory iterations, up to five local centers were allowed, subject to a maximum center-to-center spacing of ≤ 500 µm. Candidate locations violating either the tested-point exclusion zone or the safe-band constraint were automatically skipped. In addition, an edge penalty was introduced to counteract the GP's tendency to favor boundary regions toward the later stages of exploration.

Each accepted local grid center is added as a $g \times g$ grid at 5 µm pitch as tentative points for indentation. After each the GP is refit and the stay-local versus jump-global decision recomputed after every completed block. Upon completing each block of $g^2$ indents the system exports data, refreshes the training table, refits the GP, and compares staying on the next local center versus jumping to the best global candidate by evaluating uncertainty per time cost, including the 520 s reconfiguration penalty for the global move. If the local score is larger, indentation continues locally; otherwise, tests are stopped and the plan is recomputed around the new global primary. An additional cost check under slow hold evaluates whether switching to the fast hold (based on newly measured drift) would save at least the reconfiguration penalty ($g^2(80 - 5) \geq reconfiguration\ cost$); if so, tests are stopped and new global center with fast hold is selected. Figure 7a shows a 50×50 (2,500-point) grid with the x-axis reversed: in this setup, increasing stage x moves the sample right while the tip is fixed, so indentation occurs to the left. Hence x decreases toward the right. Figure 7b shows an example of all visited centers (colored by recency).

Figure 7c compares the total accumulated time versus iteration for dynamic strategy. The black curve represents a baseline GP policy that ignores reconfiguration cost and thus repeatedly jumps to the most uncertain point after each block. The orange curve shows the proposed cost-aware policy, which weighs uncertainty against execution cost, remains within local neighborhoods when efficient, and pays the 520 s reconfiguration penalty only when globally justified. Over the full run, the baseline required 150,780 s (≈42 h) to reach the same endpoint, whereas the cost-aware policy completed in 43,176 s (≈12 h). While a plain GP policy already outperforms naïve grid sampling, the cost-aware approach reduces total experiment time by roughly 3.5× through efficient local exploration and adaptive hold-time adjustment.

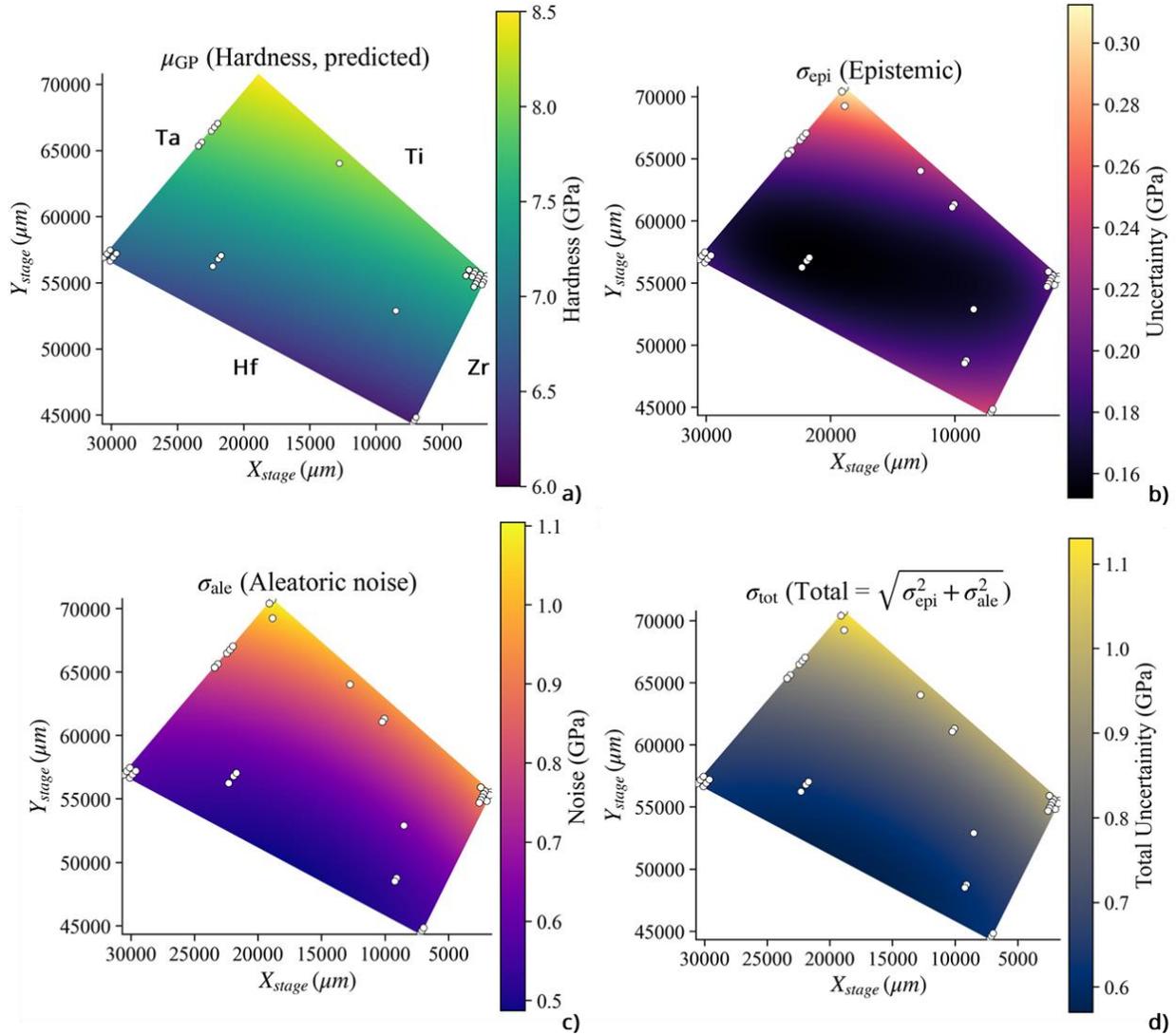

*Figure 8 (a–d) Gaussian process outputs: mean prediction, estimated noise, epistemic uncertainty, and total uncertainty. Local clusters appear where the local "stay" score exceeds the global "jump" score.*

Figure 8 summarizes the Gaussian Process (GP) prediction, noise, and uncertainty fields. Figure 8a shows the GP mean, with overlaid points marking the measurement centers. Several tight clusters correspond to locally executed $g \times g$ grids; these clusters appear when the local "stay" score (posterior uncertainty divided by local cost) exceeds the best global "jump" score (uncertainty divided by global cost plus the 520 s reconfiguration penalty). The mean surface varies smoothly across the domain, exhibiting gentle gradients from Ta/Ti-rich to Hf/Zr-rich regions—consistent with compositional variation in the quaternary system. In the absence of any change in physical mechanism, such smooth monotonic behavior is expected. The predicted hardness decreases gradually from the upper-left (Ta–Ti-rich) to the lower-right (Hf–Zr-rich) region, aligning with theoretical expectations based on previous studies)[41].

Figure 8b shows the epistemic uncertainty, derived as the standard deviation of the latent function variance returned by the heteroskedastic GP. This metric governs the selection of the next global centers. During the initial iterations, exploration naturally focuses on the corners (BL, TR, BR, TL) despite the corner penalty, as global uncertainty dominates and no region yet warrants local refinement. As observations accumulate, exploration shifts toward the interior, followed by renewed corner exploration once the algorithm finds local regions worth revisiting. The uncertainty term alone determines where to measure next, while the noise term is used to size the local grid according to the target SEM criterion.

Figure 8c shows the heteroskedastic GP's estimated noise field. The model predicts noise exceeding 10 % of the maximum hardness in some regions, varying similarly to the hardness field itself. Unlike epistemic uncertainty, the noise reflects irreducible variability arising from instrument drift, surface roughness, or film heterogeneity at the indentation scale. The noise map remains comparatively smooth and slowly varying across the domain, with slightly elevated values near the edges. Figure 8d shows the combined uncertainty field across the entire domain.

**Materials and Methods**

*Materials*

As a model system, a $Ta_wTi_xHf_yZr_z$ thin film was deposited by co-sputtering from four pure metal targets (Ta, Ti, Hf, and Zr) to fabricate a combinatorial composition library on a 100 mm-diameter fused silica substrate with a thickness of 500 µm. The deposition chamber was evacuated to a base pressure of approximately $3 \times 10^{-7}$ Torr prior to sputtering. During deposition, the sputtering powers applied to the Ta, Ti, Hf, and Zr targets were 30 W (DC), 190 W (DC), 165 W (RF), and 145 W (RF), respectively. The sputtering process was conducted for 70 minutes under a working argon pressure of 5 mTorr. After sputtering, the thin film was annealed at 1000°C for 1 hour under vacuum. The details of the composition variation and the location of the extracted subsection within the wafer are provided in the Supplementary Information.

**Conclusion**

This study establishes a cost-aware, adaptive framework for automated nanoindentation and workflow optimization that substantially accelerates materials discovery. Conventional nanoindentation workflows-whether based on dense grid searches or simple Gaussian-process exploration-are limited by the high cost of reconfiguration, drift stabilization, and redundant measurements. By explicitly modeling these costs and embedding them within a Bayesian-optimization framework, the work operationalizes a universally applicable automation strategy that accounts for instrument precision, positional repeatability, and drift behavior. The strategy combines heteroskedastic Gaussian processes, adaptive hold times, adaptive grid sizing, and meta-grid formation to simultaneously capture instrument constraints, local noise, and global uncertainty. The result is an intelligent system that balances exploration with practical execution costs, thereby improving both efficiency and data fidelity.

Through emulator studies and experimental validation on co-sputtered thin films, the proposed method demonstrates an up to thirty-fold improvement in effective throughput compared to traditional grid-based approaches. Even relative to standard Gaussian-process acquisition strategies, the incorporation of cost awareness approximately halves the total experimental time while preserving accuracy in the reconstructed hardness maps. The hierarchical meta-testing workflow, combining local-grid and global-exploration steps, prevents kernel-length collapse and ensures stable convergence across measurements with differing time scales. These gains are achieved not by compromising data quality but by strategically reducing redundant operations, adapting measurement protocols to local conditions, and minimizing high-overhead reconfigurations. The results highlight the potential of integrating decision theory with laboratory automation to overcome long-standing characterization bottlenecks in high-throughput materials research.

It is shown that the method can reduce the time taken to discover the variation of material property within the film by a factor of 10 in comparison to the GP with purely uncertainty-based acquisition function which itself has been shown to work 3-5 times faster in comparison to the standard grid search-based approaches. The methodology developed in this work is initially developed based on an emulator of a thin film. The results are experimentally confirmed using a co sputtered thin film. Therefore, the results presented are more than a proof of concept. They demonstrate that intelligent automation, when combined with practical cost considerations, can drastically accelerate materials discovery. Instead of relying on brute force mapping of the composition or process space, we leverage a principal framework that adapts to the landscape of property space and makes decisions that are both data efficient and cost effective.

The broader implication of this framework is the shift from passive, fixed protocols toward active and context-aware experimentation. By treating cost not as an afterthought but as a central component of experimental design, the workflow transforms nanoindentation into a scalable tool for combinatorial libraries, thin films, and complex microstructures. Because the emulator and cost models are architecture-agnostic, the same principles can be generalized to other autonomous or semi-autonomous instruments where positioning accuracy, stabilization time, or environmental drift constrain throughput. Beyond indentation, the general principles outlined here are applicable to a wide class of automated characterization platforms where stage movement, stabilization, or calibration times dominate the measurement budget. As synthesis methods continue to generate increasingly complex libraries, adaptive cost-aware exploration will be essential for maintaining parity between fabrication and characterization. Cost-aware Bayesian optimization coupled with automated nanoindentation provides a pathway to drastically shorten the time required to extract reliable mechanical property maps. This approach enables data-efficient exploration of large parameter spaces and offers a generalizable blueprint for accelerating materials discovery.

**Acknowledgements**
This research was primarily supported by National Science Foundation Materials Research Science and Engineering Center (MRSEC) program through the UT Knoxville Center for Advanced Materials and Manufacturing (DMR-2309083).

**Author Contributions**

V.C. and S.V.K. conceived the overall idea and research direction. V.C. performed the initial simulations, carried out the experiments, and drafted the manuscript. S.V.K. provided core conceptual guidance, contributed to the development and refinement of the framework, and completed the first round of manuscript revision. S.P. contributed to experimental validation and subsequent manuscript revisions. H.Z. and P.D.R. provided the materials used in the experimental investigation and assisted with manuscript revisions. D.P. supported the overall direction of the project and provided guidance during manuscript preparation and revision.

**Conflict of Interest**

All authors declare no financial or non-financial competing interests.

# SUPPLEMENTAL INFORMATION

## Construction of the Synthetic Hardness Emulator

A synthetic composition–property landscape was created to evaluate the automated nanoindentation workflow without relying on any specific material class. The purpose of this emulator is to generate a smooth but nontrivial property field that contains gradients, curvature, and local interactions so that the adaptive measurement algorithm can be tested in a controlled setting. The three composition fields $C_A(x,y)$, $C_B(x,y)$, and $C_C(x,y)$ were assigned spatial distributions that vary diagonally across the domain, which resembles the type of gradients produced in combinatorial sputtering. Figure S1 shows the resulting spatial maps for the three components along with the corresponding synthetic hardness field used as ground truth.

The hardness model itself is not taken from any specific alloy system. Instead, it is a phenomenological construction designed purely for generating a property landscape with realistic complexity. The first term in the model is a simple rule-of-mixtures expression that provides a baseline contribution proportional to the local compositions. The second term consists of pairwise products $C_i C_j$ scaled by interaction coefficients and is included as an artificial representation of solid-solution type interactions to introduce controlled nonlinear coupling between components. This form is not a standard description of alloy hardening but was introduced here to ensure that the hardness varies not only with individual components but also with how components coexist locally. These pairwise terms create smooth nonlinear coupling between the composition fields, which helps produce a richer landscape for testing the adaptive workflow. The final term, applied to component C, penalizes very high amounts of this component and produces a maximum at intermediate concentration. This ensures that the overall hardness field contains regions of both monotonic and non-monotonic behavior.

Figure S1 shows the three composition maps generated for the emulator and the resulting hardness surface produced by the combined effect of the rule-of-mixtures, interaction, and penalty contributions. Component A increases toward the upper left, component B toward the lower right, and component C forms a diagonal gradient across the domain. The hardness field reflects all three patterns and shows the expected smooth curvature and directional variation. This synthetic landscape served as the ground truth against which the performance of the Bayesian-optimized automated nanoindentation workflow was evaluated.

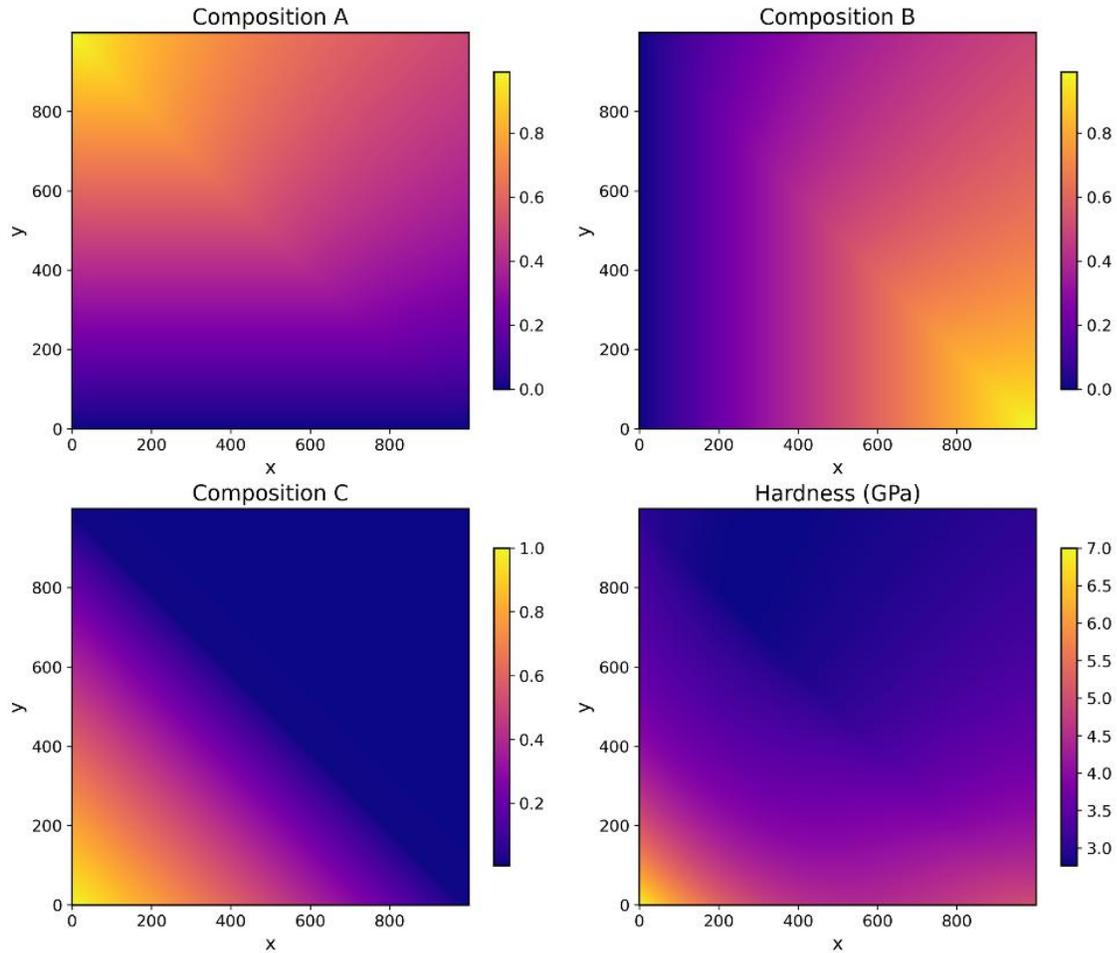

*Figure S9. Thin film emulator for a ternary system. (a, c) show the spatial variation of elements A, B and C respectively, while d) shows the corresponding hardness distribution in (x, y).*

## *Experimental thin film composition variation*

Figure S2 shows the spatial distribution of Ti, Ta, Hf, and Zr across the entire wafer and within the selected subsection used in this study. The subsection corresponds to the region bounded by $0 \leq X \leq 1.25$ and $2.34 \leq Y \leq 5$, as indicated in red. Among the four elements Ta shows the strongest variation, whereas Zr varies the least within this area.

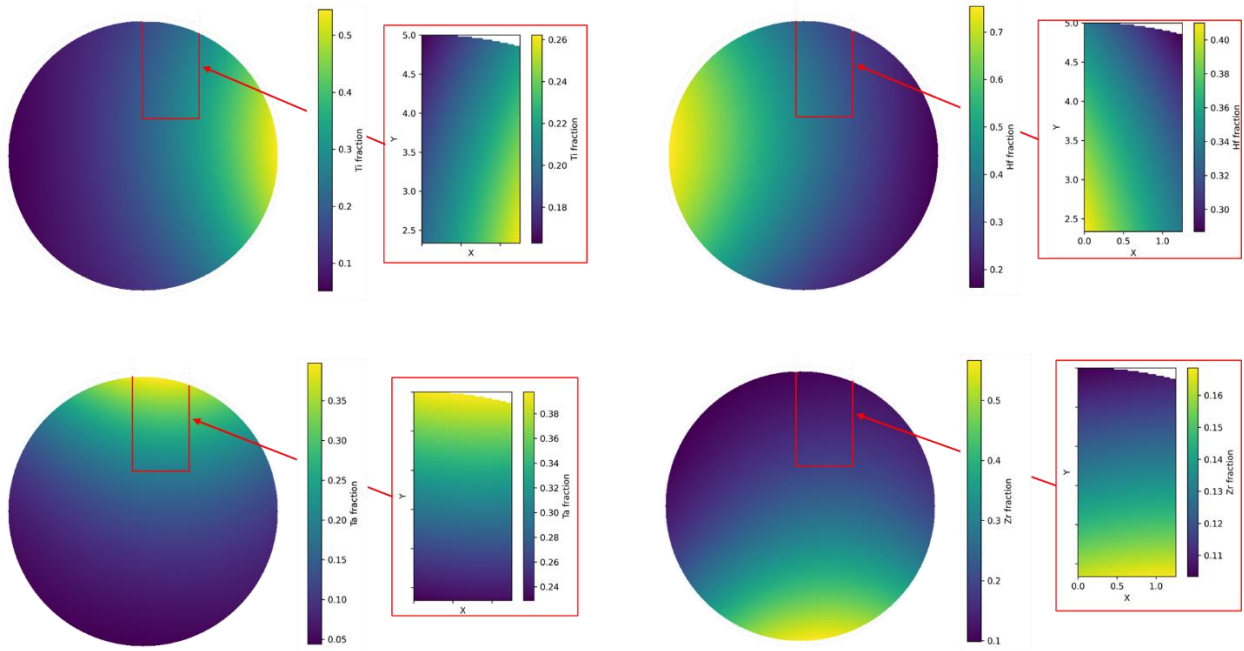

*Figure S10: Spatial distribution of Ti, Ta, Hf, and Zr within the selected subsection of the composition wafer. Each panel shows the measured elemental fraction for all points inside the defined rectangular region*